\documentclass[a4paper,14pt]{extarticle}

\usepackage{graphicx}
\usepackage{amsmath,amsthm,amssymb}

\usepackage[bindingoffset=0cm,
            left=1.5cm,
            right=1.5cm,
            top=1.5cm,
            bottom=2cm,
            footskip=1cm]{geometry}
\usepackage{float}
\usepackage{enumitem}
\usepackage{physics}
\usepackage{bm}
\usepackage{mathrsfs}  

\usepackage[dvipsnames,svgnames]{xcolor}

\usepackage{hyperref}
\hypersetup{colorlinks=true, unicode=true,menucolor=Brown,linktoc=page,
pagecolor=Brown,
citecolor=OliveGreen,linkcolor=Brown}

\title{Overdamping of \\ Neutron-Mirror-Neutron Transitions \\ in Neutron Stars \thanks{\texttt{arXiv v2 {\sc [draft 02]} (19/06/2026)}}}
\author{{\textbf{B.O. Kerbikov}}\thanks{bkerbikov@gmail.com} \smallskip \\
 Lebedev Physical Institute,\\ Moscow 119991, Russia \medskip \\
 Moscow Institute of Physics and Technology, \\ Dolgoprudny 141700, Moscow Region, Russia}

\newcommand{\beq}{\begin{eqnarray}}
 \newcommand{\eeq}{\end{eqnarray}}
\newcommand{\be}{\begin{equation}}
 \newcommand{\ee}{\end{equation}}

\def\fun#1#2{\lower3.6pt\vbox{\baselineskip0pt\lineskip.9pt
\ialign{$\mathsurround=0pt#1\hfil ##\hfil$\crcr#2\crcr\sim\crcr}}}

\newcommand{{\SD}}{\rm SD}

\newcommand{{\Mc}}{\mathcal{M}}

\begin{document}
\maketitle
\begin{abstract}
\noindent The neutron to mirror neutron transitions in neutron stars would possibly result in significant effects. In this work we show that collisional decoherence entails exponential relaxation in lieu of oscillations. Decoherence is a great many orders of magnitude faster than the expected oscillations. The admixture of mirror neutrons at all times remains very small with respect to ordinary neutrons component.
\end{abstract}

\section{Introduction}

The idea of the existence of a hypothetical hidden mirror sector of the Standard model was first explicitly formulated in \cite{01}. It took a distinct form in \cite{02}. The subject has a long history -- for the review of the emergence and evolution of the concept see \cite{03}. 

Possible existence of mirror matter (MM) and the intersection of MM and the ordinary matter (OM) have been the subject of intense studies during the last decades. The significance of MM for cosmology, astrophysics, and dark matter search has been discussed in a number of papers \cite{04,05,06,07,08,09,10,11}. The MM and OM interact via gravity. Besides gravity, other tiny cross-interactions between the two sectors possibly exist and can give rise to interesting observable effects.

In particular, the neutron $n$ can be mixed with the mirror neutron $n'$ leading to the $n$-$n'$ transitions as it was suggested in \cite{12,13}.

Since then a number of phenomenological applications of $n$-$n'$ oscillations have been discussed. It includes the Greisen--Zatsepin--Kuzmin cutoff \cite{12,13}, suppression or enhancement of oscillations due to non-zero mirror magnetic field $B'$ and its interplay with the ordinary field $B$ \cite{14}, the role of $n$-$n'$ transitions in the neutron lifetime puzzle \cite{15}. An important question was raised already in~[12,13], namely whether $n-n'$ transitions can proceed in nuclei. The short answer given in \cite{12,13} was that they are forbidden by the neutron binding energy as far as $n$ and $n'$ are degenerate in mass ($Z_2$ symmetry). According to \cite{13} similar arguments apply to the neutron stars (NSs). The last statement underwent an important evolution \cite{16,17,18}. The $n$-$n'$ conversion in NS is allowed since the binding of neutrons in NS is provided mainly by the gravitational pressure \cite{18}. The hole created by the neutron conversion is filled by the neighbouring $n$ while the created $n'$ gravitates to the NS center. In view of the breakthrough in observations of NSs \cite{19} the problem of NS transformation into a partially consisting of MM mixed star (MS) is now under intense discussion \cite{18,20,21,22,23,24}. The rate of this process and the observable manifestations of the MS formation are addressed in these publications.



In the present work we concentrate on the renormalization of the $n$-$n'$ transition inside the NS caused by the incoherent collisions with the surrounding neutrons. The point is that the rate of oscillations critically depends on the ambient conditions which include the surrounding medium, the external magnetic field $B$, and mirror magnetic field $B'$. We show that the coherence needed for $n$-$n'$ transition to occur is ruined by high frequency neutron collisions in the NS matter. A correct way to describe the quantum system in contact with the environment is to use the density matrix formalism \cite{25,26} and the theory of open quantum systems \cite{27,28} based on the Lindblad \cite{29,30} and Bloch \cite{31,32} equations. Put differently, the open quantum system can not be described by the two-by-two Hamiltonian not unless one is at risk to obtain dubious results. The problem of oscillations in a two-state physical system interacting with the environment was for the first time formulated and solved using the density matrix in a seminal paper by G. Feinberg and S. Weinberg \cite{33} devoted to muonium to antimuonium conversion. A clear and comprehensive example of applying the above formalism to treat positronium to mirror positronium oscillations is presented in \cite{34}.

The paper is structured as follows. In Sec. II we introduce the density matrix formalism and Bloch equation. As a warming up exercise the standard equation for the $n$-$n'$ oscillations in vacuum is derived. In Sec. III the Lindblad and Bloch equations are introduced and discussed. In Sec. IV we apply Lindblad and Bloch equations to $n$-$n'$ transition inside the NS and derive the overdamped relaxation in lieu of oscillations. In Sec. V we discuss the intricate interplay of the collisions friction parameter and the time interval between collisions. Conclusions and future research perspectives are presented in Sec. VI.

\section{The density matrix approach to $n$-$n'$ oscillations in vacuum} 

As a warming-up task we solve the problem of $n$-$n'$ oscillations in vacuum by means of
the density matrix formalism. For $n$-$n'$ system the density matrix reads (the indices
$1$ and $2$ are attributed to $n$ and $n'$)
\begin{equation}
\hat{\rho}(t)=
\begin{pmatrix}
\psi_1\psi_1^{*} & \psi_1\psi_2^{*}\\
\psi_2\psi_1^{*} & \psi_2\psi_2^{*}
\end{pmatrix}
=
\begin{pmatrix}
\rho_{11} & \rho_{12}\\
\rho_{21} & \rho_{22}
\end{pmatrix}
\label{1}
\end{equation}

Consider an isolated $n$-$n'$ system in vacuum with $\beta$-decay neglected.
We assume that the mirror $Z_2$ symmetry is exact so that the fermion masses in both sectors
are the same. The Hamiltonian has the form
\begin{equation}
\hat{H}=
\begin{pmatrix}
E & \varepsilon\\
\varepsilon & E'
\end{pmatrix},
\label{2}
\end{equation}
where $\varepsilon$ is a mixing parameter and $E-E' = d$.
Currently we do not specify the origin and the value of the parameter $d$.
If we identify it with the energy difference between $n$ and $n'$ due to the superimposed magnetic field $B$ in absence of mirror field $B'$, then $d = |\mu_n B| = 6 \cdot 10^{-12}
\left( \frac{B}{1\,\mathrm{G}} \right)\,\mathrm{eV}.$
Laboratory experiments on $n$-$n'$ are typically done for $B < 1\,\mathrm{G}$.
For magnetars $ B \simeq (10^{12}-10^{14})\,\mathrm{G}. $
For the mixing parameter we may take the value
\[
\varepsilon < 1.5 \cdot 10^{-18}\,\mathrm{eV}
\]
from the direct experiment \cite{35,36}. It corresponds to the oscillation time $\tau_{nn'} > 448\,\mathrm{s} $ \cite{35}. This value was obtained under the assumption of zero mirror magnetic field and perfect degeneracy of $n$ and $n'$ masses.

Dropping the assumption $B' = 0$ leads to the dependence of the lower limit on $\tau_{nn'}$ on the strength and the angle between $B$ and $B'$ \cite{14,37}. The search for $n$-$n'$ oscillations with the account of possible mirror magnetic field has been performed in \cite{38} with the result $\tau_{nn'} > (6-9)\,\mathrm{s}$ for $ B' \simeq (10-20)\,\mu\mathrm{T}.$ Clear indications on the signal-like anomalies in favor of $B'$ manifestation have not been observed. Under the assumption $B' = 0$ this experiment yielded the result $\tau_{nn'} > 352\,\mathrm{s}$ $\left( \varepsilon < 1.9 \cdot 10^{-18}\,\mathrm{eV} \right).$ Note that the values of $\varepsilon$ as large as $10^{-15}\,\mathrm{eV}$ \cite{12,18} and even $\varepsilon \sim 10^{-13}\,\mathrm{eV}$ \cite{20} are discussed in the literature. Note that $\varepsilon$ sharply depends on the mass splitting $\Delta m = m - m'.$ We remind that in this article we consider the perfectly degenerate masses of $n$ and $n'$. Otherwise $\Delta m$ might have contributed to $d$. The detailed investigation of $\Delta m \neq 0$ case is given in \cite{39}. We stick to the exact $Z_2$ symmetry in order not to make less clear our core message and not to introduce additional complications. The above estimates and considerations allow to conclude that $d \gg \varepsilon$ is a reliable conjecture.


The time evolution of $\hat{\rho}(t)$ is described by the Von-Neumann--Liouville equation
\begin{equation}
\frac{d\hat{\rho}}{dt}=-i\,[\hat{H},\hat{\rho}].
\label{3}
\end{equation}

It yields four coupled linear differential equations
\begin{align}
\dot{\rho}_{11} &= -i\varepsilon(\rho_{21}-\rho_{12}), \label{4}\\
\dot{\rho}_{12} &= i\varepsilon(\rho_{11}-\rho_{22})-i d\,\rho_{12}, \label{5}\\
\dot{\rho}_{21} &= -i\varepsilon(\rho_{11}-\rho_{22})+i d\,\rho_{21}, \label{6}\\
\dot{\rho}_{22} &= i\varepsilon(\rho_{21}-\rho_{12}). \label{7}
\end{align}

The quantity of interest $\rho_{22}(t)=\abs{\psi_{n'}(t)}^2$ may be found directly by
solving this set of equations. For further purposes it makes sense to act differently,
namely to reformulate the Von-Neumann--Liouville equation in terms of Bloch 3-vector
$\bm{R}$ \cite{31,32}. It is introduced by the expansion of the density matrix over the Pauli
matrices
\begin{equation}
\hat{\rho}=\frac{1}{2}\left(1+\bm{R}\cdot\bm{\sigma}\right),
\label{8}
\end{equation}
\begin{equation}
\bm{R}=
\begin{pmatrix}
\rho_{12}+\rho_{21}\\
-i(\rho_{21}-\rho_{12})\\
\rho_{11}-\rho_{22}
\end{pmatrix}.
\label{9}
\end{equation}

The Von-Neumann--Liouville equation (\ref{3}) is equivalent to the equation of motion of the
Bloch vector
\begin{equation}
\dot{\bm{R}}=\bm{V}\times \bm{R},
\label{10}
\end{equation}
where
\begin{equation}
\bm{V}=
\begin{pmatrix}
2\varepsilon\\
0\\
d
\end{pmatrix}.
\label{11}
\end{equation}

This form of Bloch equation was proposed by Leo Stodolsky \cite{28}. Equation (\ref{10}) describes
the precession of the Bloch vector $\bm{R}$ around the ``magnetic field'' $\bm{V}$.
According to (\ref{10}) the length of $\bm{R}$ does not change. This means the absence of
decoherence. Decoherence does not happen as soon as the system is isolated from the
environment. In terms of $\bm{R}$ the system of equations (\ref{4})-(\ref{7}) reduces to the following
\begin{equation}
\dot{R}_x=-d R_y,\qquad
\dot{R}_y=d R_x-2\varepsilon R_z,\qquad
\dot{R}_z=2\varepsilon R_y.
\label{12}
\end{equation}

We solve these equations with the initial condition $\bm{R}(t=0)=(0,0,1)$ which means that
the system is initially in the $\ket{n}$ state. Taking in (\ref{12}) the second time derivative
of $R_y$ and using the other two equations one arrives at the result
\begin{equation}
\ddot{R}_y+\mathscr{D}^2 R_y=0,
\label{13}
\end{equation}
where
\begin{equation}
\mathscr{D}^2=d^2+\varepsilon^2.
\label{14}
\end{equation}

Insertion of the solution of (\ref{13}) into the third equation (\ref{12}) yields
\begin{equation}
R_z(t)=\rho_{11}-\rho_{22}
=1-\frac{8\varepsilon^2}{\mathscr{D}^2}\sin^2\frac{\mathscr{D}}{2}t.
\label{15}
\end{equation}

Invoking the condition $\rho_{11}+\rho_{22}=1$, one obtains the well-known result
\begin{equation}
\rho_{22}(t)=\abs{\psi_{n'}(t)}^2
=\frac{4\varepsilon^2}{\mathscr{D}^2}\sin^2\frac{\mathscr{D}}{2}t.
\label{16}
\end{equation}

In the ultra-short time limit $\mathscr{D} t \ll 1$ (\ref{16}) yields
\begin{equation}
\rho_{22}(t)\approx \varepsilon^{2} t^{2}.
\label{17}
\end{equation}

This law does not allow to define the transition probability per unit time \cite{25}.
The oscillation formula (\ref{16}) could have been obtained by solving the Schrodinger equation.
Our purpose was to introduce the density matrix formalism which lies at the core of the \emph{correct approach}
to quantum systems in contact with the environment. The problem of $n$-$n'$ oscillations in the framework of the density matrix formalism with absorption has been considered in \cite{40}.

\section{Lindblad and Bloch Evolution Equations}

The departure from the ordinary quantum mechanics is needed to describe a system that interacts with the surroundings.
The Lindblad and Bloch equations are the instruments suited for this purpose.
The $n$-$n'$ system in a NS may be regarded as a part of a large system composed of the $n$-$n'$ subsystem and the ensemble
of neutrons forming the NS. The density matrix of the full system evolves according to the
Von--Neumann--Liouville equation (\ref{3}). The reduced density matrix \cite{26,27} of the $n$-$n'$ subsystem
is obtained by the partial trace over the unobserved states of the environment \cite{26}
\begin{equation}
\hat{\rho}(t)=\mathrm{tr}_{\mathrm{env}}\!\left(\hat{\rho}_{\mathrm{full}}(t)\right).
\label{18}
\end{equation}

The evolution of the reduced density matrix is described by the Lindblad equation \cite{29,30}
\begin{equation}
\frac{d\hat{\rho}}{dt}
= -i\,[\hat{H},\hat{\rho}]
+ \sum_{n}\left[\hat{L}_{n}\hat{\rho}\hat{L}_{n}^{\dagger}
-\frac{1}{2}\left\{\hat{L}_{n}^{\dagger}\hat{L}_{n},\hat{\rho}\right\}\right].
\label{19}
\end{equation}

The first term in (\ref{19}) is the usual Schrodinger term, $\hat{L}_n$ is a set of additional operators called Lindblad or jump operators, $\{\cdot,\cdot\}$ is the anticommutator.
The second term in (\ref{19}) is the dissipative one responsible for the loss of information into the environment.
It transforms pure states into mixed ones inducing decoherence.
A lucid pedagogical derivation of the Lindblad equation may be found in \cite{41}.
This equation has been used to describe oscillations in a wide range of physical systems from the atomic clock \cite{42}
and ultracold atoms \cite{43} to oscillating neutrino scattering on leptons in plasma \cite{44},
heavy quark dynamics in quark-gluon plasma \cite{45}, bottomonium regeneration \cite{46}, Hawking radiation \cite{47}.

\section{Neutron--Mirror-Neutron Conversion in a Neutron Star}

First we formulate the underlying assumptions under which we consider the $n$-$n'$ conversion in the NS.
They include:
\begin{enumerate}
\item $Z_2$ mirror symmetry is exact so that $n$ and $n'$ have equal masses
\item mirror neutron does not scatter of the ordinary matter (OM) neutrons forming the NS and there is no $n'$-$n$ regeneration,
\item the admixture of MM in NS which may cause $n'$-$n'$ scattering is neglected,
\item $\beta$-decay in both $n$ and $n'$ channels is temporary omitted.
\end{enumerate}

Some comments on the above assumptions are needed. There are arguments to soften the exact $Z_2$ symmetry. Big Bang Nucleosynthesis requires the asymmetric inflationary reheating in the two sectors \cite{06,48}. In addition, breaking of $Z_2$ may serve as an explanation of the neutron lifetime puzzle \cite{15}. As mentioned previously, in order to avoid complications unrelated to the topic of this work we assume the perfect $n$ and $n'$ mass degeneracy. The next remark is that by NS number density we imply the neutron density leaving aside the admixture of protons and hyperons. The $n'$-$n$ regeneration is blocked by Pauli principle. For demonstration purposes we choose the number density twice the value of the normal nuclear density keeping in mind the density variation from the crust to the core.

Under the listed assumptions the elements entering into the Lindblad equation (\ref{19}) are of the following form
\begin{equation}
\hat{H}=
\begin{pmatrix}
E-\dfrac{2\pi}{k}\,n v\,\mathrm{Re}\,f(0) & \varepsilon\\[6pt]
\varepsilon & E'
\end{pmatrix},
\label{20}
\end{equation}
\begin{equation}
\hat{L}=\sqrt{n v}\ \hat{F},
\qquad
\hat{F}=
\begin{pmatrix}
f(\theta) & 0\\
0 & 0
\end{pmatrix}.
\label{21}
\end{equation}

Here $k$ is the neutron momentum, $n$ is the NS number density, $v$ is the neutron velocity,
$f(\theta)$ is the neutron-neutron elastic scattering amplitude.
The term $\dfrac{2\pi}{k}\,n v\,\mathrm{Re}\,f(0)$ corresponds to the energy shift due to forward scattering.
Inserting (\ref{20}) and (\ref{21}) into (\ref{19}) one arrives after a simple algebra at a set of four coupled differential equations

\begin{align}
\dot{\rho}_{11}=&-i \varepsilon\left(\rho_{21}-\rho_{12}\right) \label{22}\\
\dot{\rho}_{22}=&i \varepsilon\left(\rho_{21}-\rho_{12}\right) \label{23}\\
\dot{\rho}_{12}=&-M \rho_{12}-i(d+K) \rho_{12}+i \varepsilon\left(\rho_{11}-\rho_{22}\right) \label{24}\\
\dot{\rho}_{21}=&-M \rho_{21}+i(d+K) \rho_{21}-i \varepsilon\left(\rho_{11}-\rho_{22}\right) \label{25}
\end{align}

The quantities $M$ and $K$ stand for
\begin{equation}
M=\frac{2\pi}{k}\,n\,v\,\Im f(0), \qquad
K=-\frac{2\pi}{k}\,n\,v\,\Re f(0). \label{26}
\end{equation}

According to the optical theorem $M=\frac{1}{2}\,n\sigma v$ with
$\sigma$ being the $nn$ cross section, $K$ gives the energy due to the
index of refraction. The coefficient $\frac{1}{2}$ in $M$ comes from the active component which can scatter \cite{28,33,34}. To our knowledge, the set of equations (\ref{22})-(\ref{25})
do not have a transparent analytical solution.

Equations (\ref{22})-(\ref{25}) are equivalent to the following matrix equation for
the Bloch vectors $\vec{R}$
\begin{equation}
\dot{\vec{R}}=\vec{V}\times\vec{R} - D_T\,{R}_T, \label{27}
\end{equation}

where
\begin{equation}
\vec{V}=
\begin{pmatrix}
2\varepsilon\\
0\\
d+K
\end{pmatrix},
\qquad
D_T=
\begin{pmatrix}
M & 0\\
0 & M
\end{pmatrix},
\qquad
{R}_T=
\begin{pmatrix}
R_x\\
R_y
\end{pmatrix}.
\label{28}
\end{equation}

The difference of (\ref{27})-(\ref{28}) from (\ref{10})-(\ref{11}) is the presence of the term
$D_T\,{R}_T$ which represents the quantum friction and is responsible
for the loss of coherence. It leads to the destruction of the
off-diagonal elements of the density matrix.

Likewise the Lindblad equations (\ref{22})-(\ref{25}) the Bloch equations (\ref{27})-(\ref{28})
can hardly be solved analytically. To proceed further we drop the
component $V_z=(d+K)$ of $\vec{V}$. It corresponds to the contributions
of the external magnetic field and the index of refraction. The component
$V_z$ does not participate in decoherence. With $V_z$ omitted the
evolution equations for $R_z$ and $R_y$ are

\begin{align}
\dot{R}_z &= 2\varepsilon\,R_y, \notag \\
\dot{R}_y &= -M R_y - 2\varepsilon\,R_z. \label{29}
\end{align}

Taking in (29) the second time derivative of $R_z$ one arrives at the equation
\begin{equation}
\ddot{R}_z + M\dot{R}_z + 4\varepsilon^2 R_z = 0. \label{30}
\end{equation}

This is the equation of oscillator with friction $M$. It possesses three
types of solutions depending on the relative magnitude of the damping
parameter $M$ and the tunneling parameter $\varepsilon$. In the next
Section we shall relate $M$ to the flight-time $t_{nn}$ between collisions.
The connection is not entirely simple as might be expected. The character
of the solution depends on the sign and the value of the quantity
\begin{equation}
\Omega^2=\frac{M^2}{4}-4\varepsilon^2. \label{31}
\end{equation}

If $\Omega^2>0$ the interaction with the environment destroys the off-diagonal elements of the density matrix. This is the
overdamping regime. The opposite underdamped case $\Omega^2<0$
corresponds to slightly or strongly suppressed oscillations. The critical
damping $\Omega^2=0$ implies a fine tuning of $M$ and $\varepsilon$.
Finally, for $M=0$ the system is in free oscillation regime.

It is not difficult to prove that $n$-$n'$ conversion in the NS proceeds
in the overdamped mode. To be aware of this we compare the values of $M$
and $\varepsilon$. For the value of $\varepsilon$ we set $\varepsilon = 1.5\cdot 10^{-18}\,\text{eV} \simeq 2\cdot 10^{-3}\,\text{s}^{-1}$ from the direct experiment \cite{35}. As we mentioned in Sec.~II a wide
interval of $\varepsilon$ values is discussed in literature. To estimate $M$
we proceed as follows. We take $n = 2n_0 = 0.34\,\text{fm}^{-3}$ for the number density. There is a scarce
information regarding the $nn$ collision cross section in NS. We employ the
data plotted in Fig.~7 of \cite{49}. At the above density $\sigma \simeq 30\,\text{mb}
= 3\,\text{fm}^2$. The authors of \cite{49} emphasize that the cross section in
medium is not a well defined quantity and hence the above number should be
taken with caution. The neutron velocity may be estimated as
$v \simeq \dfrac{k_F}{m} \simeq \dfrac{(3\pi^2 n)^{1/3}}{m} \simeq 0.4.$

With the above numbers one gets the following estimate for $M$
\begin{equation}
M=\frac{1}{2}n\sigma v \simeq 0.4\cdot 10^{8}\,\text{eV}
=0.6\cdot 10^{23}\,\text{s}^{-1}. \label{32}
\end{equation}

These crude estimates confirm the conclusion that $M\gg \varepsilon$,
namely $M/\varepsilon \sim 10^{25}-10^{26}$.

The solution of Eq.~(\ref{30}) with the initial condition $R_{z}(0)=1$ (pure $|n\rangle$ state) is given by
\begin{equation}
R_{z}(t)= e^{-\frac{M t}{2}}
\left[
\frac{\frac{M}{2}+\Omega}{2\Omega}e^{\Omega t}
+\frac{\frac{M}{2}-\Omega}{2\Omega}e^{-\Omega t}
\right].
\label{33}
\end{equation}

The two terms in Eq.~(\ref{33}) behave as $\exp(-\omega_i t)$, $i=1,2$ with dissimilar frequencies
\begin{equation}
\omega_1=\frac{M}{2}-\Omega \simeq \frac{4\varepsilon^2}{M},
\qquad
\omega_2=\frac{M}{2}+\Omega \simeq M.
\label{34}
\end{equation}

Note that $\omega_1/\omega_2 \simeq 4\varepsilon^2/M^2 \ll 1$.
It is clear that the large-time ($t\gg 1/M$) behavior is dominated by the first term with the eigenfrequency
$\omega_1=4\varepsilon^2/M$.
The second term can only be probed at short times $t\ll 1/M$.
Interference effects at intermediate times is a subject of a paper in preparation.
At ``long time'' $R_z(t)$ becomes an overdamped solution describing
the relaxation without oscillations with constant decoherence rate
\begin{equation}
R_{z}(t)\sim \exp\!\left(-\frac{4\varepsilon^{2}}{M}\,t\right).
\label{35}
\end{equation}

This form of $R_{z}(t)$ is in complete agreement with the corresponding results of \cite{28} and \cite{34}.
Invoking that $R_z=\rho_1-\rho_2$ one can split $R_z(t)$ into two terms
\begin{equation}
R_{z}(t)=\rho_1-\rho_2 \simeq
\exp\!\left(-\frac{4\varepsilon^{2}}{M}\,t\right)
-\frac{4\varepsilon^{2}}{M^{2}}\exp\!\left(-\frac{4\varepsilon^{2}}{M}\,t\right).
\label{36}
\end{equation}

This expression is becoming exact for $t\gg 1/M$.
The first term in (\ref{36}) describes the evolution of $|\psi_n(t)|^2$.
The disappearance time $T=M/4\varepsilon^2$ gets tremendously prolonged due to high frequency collisions inside the NS.
The mirror neutron component $|\psi_{n'}(t)|^2$ is given by the second term and is damped by a huge factor
$4\varepsilon^2/M^2$ at all times.
Equation similar to (\ref{36}) has been derived in \cite{50} for the
the neutron-antineutron conversion in nuclei.
The same type of equations describe the probability of a photon oscillating into a hidden photon in medium \cite{51,52}.
According to (\ref{36}) the $n$-$n'$ transition rate $\Gamma(n \text{-} n')$ is equal to
\begin{equation}
\Gamma(n \text{-} n')=\frac{4\varepsilon^{2}}{M} \simeq 10^{-20} \text{ yr.}^{-1} \simeq 0.2 \cdot 10^{-51} \text{ GeV} \text{ for } \varepsilon = 1.5 \cdot 10^{-18} \text{ eV}.
\label{37}
\end{equation}

Pay attention that the quantity $M$ did not come from the two-by-two \mbox{Hamiltonian}.
It originates from the Lindblad operator and is responsible for the decoherence.
In what follows we shall discuss the relation between $M$ and the time interval $t_{nn}$ between collisions.

\section{Implications}

We have shown that $n$-$n'$ oscillations in NSs are overdamped due to the difference in scale between the mixing
parameter $\varepsilon$ and the collision friction parameter $M$.
The $n$-$n'$ system evolves slowly compared to the timescales of its interaction with the environment.
It is desirable to obtain the relation between the damping factor $M$ and the time interval $t_{nn}$ between collisions.
We have estimated $M$ as $M\simeq 10^{23}\ \mathrm{s}^{-1}$  (\ref{32}). On dimensional grounds it is clear that $t_{nn}\sim 1/M$.
It turns out that to find the proportionality coefficient is not a trivial matter.
Recall that according to (\ref{26}) $M=\frac{1}{2}\,n\,\sigma\,v$.
To relate $M$ and $t_{nn}$ it is tempting to bring in the simplest
expression $l = 1/(n\sigma)$ for the neutron mean free path (MFP) and to obtain
$\Gamma(n \text{-} n')=8\varepsilon^2 t_{nn}$ replacing (\ref{37}).
The result, namely the coefficient $8$, can hardly be considered reliable.
The evaluation of MFP in dense interacting matter is a complicated problem involving the whole machinery of many-body
theory. The encountered difficulties include the choice of the potential, the account of the Pauli principle, the
three body forces, the geometry of the trajectory, etc.
Despite the great efforts of several authors in various approaches (see a list of references in \cite{49}) the problem is
still pending.
Therefore the expression $l=1/(n\sigma)$ as well as $\Gamma_{nn}=\sigma_{nn} v n=1/t_{nn}$ used in \cite{20}
is merely correct on the dimensional grounds. Corrections induced by Pauli blocking and medium-dependent energy splitting between $n$ and $n'$ have been proposed in \cite{18,21}. However, the approach based on the Schrodinger equation with two-by-two matrix Hamiltonian cannot describe the damping of the $n$-$n^{\prime}$ conversion caused by the collisional decoherence. The Lindblad term (\ref{19}) must be added \cite{27,28,29,30,42}. The relaxation pattern (\ref{35})-(\ref{36}) sets in instead of oscillations. The transition rate in given by (\ref{37}) or by $\Gamma\left(n\text{-}n^{\prime}\right) \sim \varepsilon^2 t_{nn}$ with the proportionality coefficient and $t_{nn}$ can at best be estimated in a model-dependent way.

\section{Conclusions and outlook}

In this paper a new approach to the process of Neutron-Mirror-Neutron transition in NS is proposed. The research is driven by the idea that high frequency collisions with the surrounding neutrons cause a deep change in the way the $n$-$n'$ conversion proceeds. The $n$-$n'$ can be regarded as a subsystem in contact with the environment. Its time evolution is described as the equation of motion of the reduced density matrix $\hat{\rho}(t)$. It is no longer the Liouville--von Neumann one but contains the new terms in the form of Lindblad operators. These terms are responsible for the interaction with the environment. Under their action the off-diagonal elements of $\hat{\rho}(t)$ decay to zero. This process is referred to as decoherence or density matrix collapse. The Lindblad equation may be recasted into Bloch equation for the 3-vector $\vec{R}$ on the Bloch sphere. The equation for $\vec{R}$ is the equation for the oscillator with damping parameter $M$ which is proportional to the rate of neutron collisions in the NS. For the NS density this rate is $\sim 25$ orders of magnitude higher than the predicted $n$-$n'$ oscillation rate. Overdamping with no oscillations at all times takes place. 

For ``long'' times $t\gtrsim 10^{-23}\ \mathrm{s}$ the time dependence of ordinary and mirror components reads
\begin{equation}
\rho_{11}=|\psi_n(t)|^{2}\simeq \exp\!\left(-\frac{4\varepsilon^{2}}{M}t\right),\quad
\rho_{22}=|\psi_{n'}(t)|^{2}\simeq \frac{4\varepsilon^{2}}{M^{2}}\exp\!\left(-\frac{4\varepsilon^{2}}{M}t\right).
\label{38}
\end{equation}

In order to compare with the results of other authors it is instructive, following \cite{18,53}, to express the mixing parameter $\varepsilon$ in terms of $\varepsilon_{15} = \frac{\varepsilon}{10^{-15}\,\mathrm{eV}}.$ For $\varepsilon = 1.5 \cdot 10^{-18}\,\mathrm{eV}$ one has $\varepsilon_{15}^{2} = 2.25 \cdot 10^{-6}.$ With this value of $\varepsilon_{15}^{2}$ and the ratio $\xi = 2$ of the NS density to normal nuclear density $n_{0} = 0.16\,\mathrm{fm}^{-3},$ Eq.~(40) of \cite{18} for the effective ``starting'' rate of the NS transformation yields $\Gamma(n\text{-}n') \simeq 0.4 \cdot 10^{-52}\,\mathrm{GeV},$ and Eq.~(18) of \cite{53} gives $\Gamma(n\text{-}n') \simeq 0.2 \cdot 10^{-52}\,\mathrm{GeV}$ (in the last case the NS density is the average over the NS profile).

Note that about an order of magnitude difference between the results of \cite{18,53} and Eq.~(\ref{37}) is not very significant in view that all three results for $\Gamma(n\text{-}n')$ predict the conversion time larger than the universe age. What differs our result from that of \cite{18,53} is that we predict the MM admixture to be small at all times while according to \cite{18,53} the NS evolution ends up with equal amounts of OM and MM. And we do not need to say that the theoretical approaches of \cite{18,53} and the present one are different.

\noindent 
\section{Acknowlegments}

The author is indebted to Yu.\,A.\,Kamyshkov, M.\,I.\,Krivoruchenko and departed Iosif Khriplovich for discussions and to M.\,S.\,Lukashov for collaboration in preparing the article.


\begin{thebibliography}{99}

\bibitem{01} 
T. D. Lee and C. N. Yang, \textit{Phys. Rev.} \textbf{104}, 254 (1956).

\bibitem{02} 
I. Kobzarev, L. Okun, and I. Pomeranchuk, \textit{Sov. J. Nucl. Phys.} \textbf{3}, 837 (1966).

\bibitem{03} 
L. B. Okun, \textit{Phys. Usp.} \textbf{50}, 380 (2007), \texttt{arXiv:hep-ph/0606202}.

\bibitem{04} 
S. Blinnikov and M. Khlopov, \textit{Sov. J. Nucl. Phys.} \textbf{36}, 472 (1982).

\bibitem{05} 
H. M. Hodges, \textit{Phys. Rev.} \textbf{D 47}, 456 (1993).

\bibitem{06} 
Z. G. Berezhiani, A. D. Dolgov, and R. N. Mohapatra, \textit{Phys. Lett.} \textbf{B 375}, 26 (1996), \texttt{arXiv:hep-ph/9511221}.

\bibitem{07} 
R. N. Mohapatra, S. Nussinov, and V. L. Teplitz, \textit{Phys. Rev.} \textbf{D 66}, 063002 (2002), \texttt{arXiv:hep-ph/0111381}.

\bibitem{08} 
R. Foot, \textit{Int. J. Mod. Phys.} \textbf{A 29}, 1430013 (2014), \texttt{arXiv:1401.3965 [astro-ph.CO]}.

\bibitem{09} 
S. I. Blinnikov, ``A quest for weak objects and for invisible stars'', \texttt{arXiv: astro-ph/9801015}.

\bibitem{10} 
R. Foot, \textit{Phys. Lett.} \textbf{B 452}, 83 (1999), \texttt{arXiv:astro-ph/9902065}.

\bibitem{11} 
R. N. Mohapatra and V. L. Teplitz, \textit{Phys. Lett.} \textbf{B 462}, 302 (1999), \texttt{arXiv:astro-ph/9902085}.

\bibitem{12}
Z. Berezhiani and L. Bento, \textit{Phys. Rev. Lett.} \textbf{96}, 081801 (2006), \texttt{arXiv:hep-ph/0507031}.

\bibitem{13}
Z. Berezhiani and L. Bento, \textit{Phys. Lett.} \textbf{B 635}, 253 (2006), \texttt{arXiv:hep-ph/0602227}.

\bibitem{14}
Z. Berezhiani, \textit{Eur. Phys. J.} \textbf{C 64}, 421 (2009), \texttt{arXiv:0804.2088 [hep-ph]}.

\bibitem{15}
Z. Berezhiani, \textit{Eur. Phys. J.} \textbf{C 79}, 484 (2019), \texttt{arXiv:1807.07906 [hep-ph]}.

\bibitem{16} 
Z. Berezhiani, ``Unusual effects in $n$-$n'$ conversion'', Talk at the Workshop INT-17-69W, Seattle, 23--27 Oct. 2017.

\bibitem{17}
Z. Berezhiani, \textit{Lett. High Energy Phys.} \textbf{2}, 118 (2019), \texttt{arXiv:1812.11089 [hep-ph]}.

\bibitem{18} Z. Berezhiani, R. Biondi, M. Mannarelli, et al., \textit{Eur. Phys. J.} \textbf{C 81}, 1036 (2021), \texttt{arXiv:2012.15233 [astro-ph.HE]}.

\bibitem{19} 
J. M. Lattimer, \textit{Ann. Rev. Nucl. Part. Sci.} \textbf{71}, 433 (2021).

\bibitem{20} 
I. Goldman, R. N. Mohapatra, and S. Nussinov, \textit{Phys. Rev.} \textbf{D 100}, 123021 (2019), \texttt{arXiv:1901.07077 [hep-ph]}.

\bibitem{21} 
D. McKeen, M. Pospelov, and N. Raj, \textit{Phys. Rev. Lett.} \textbf{127}, 061805 (2021), \texttt{arXiv:2105.09951 [hep-ph]}.

\bibitem{22} 
B. O. Kerbikov, \textit{Phys. Rev.} \textbf{D 106}, 015015 (2022), \texttt{arXiv:2112.14157 [hep-ph]}.

\bibitem{23} 
I. Goldman, R. N. Mohapatra, S. Nussinov, and Y. Zhang, \textit{Phys. Rev. Lett.} \textbf{129}, 061103 (2022), \texttt{arXiv:2208.03771 [hep-ph]}.

\bibitem{24} 
I. Goldman, R. N. Mohapatra, S. Nussinov, and Y. Zhang, \textit{Eur. Phys. J.} \textbf{C 82}, 945 (2022), \texttt{arXiv:2203.08473 [hep-ph]}.

\bibitem{25} 
L. D. Landau and E. M. Lifshitz, \textit{Quantum Mechanics: Course of Theoretical Physics}, Vol.~3 (Pergamon Press, Oxford, 1978).

\bibitem{26} 
R. P. Feynman, \textit{Statistical Mechanics: A Set of Lectures} (W. A. Benjamin Inc., Mass., 1972).

\bibitem{27} 
H.-P. Breuer and F. Petruccione, \textit{The Theory of Open Quantum Systems} (OUP, New York, 2002).

\bibitem{28} 
L. Stodolsky, ``Quantum Damping and its Paradoxes'' in \textit{Quantum Coherence}, Ed. by J. S. Anandan (World Scientific, Singapore, 1990).

\bibitem{29} 
G. Lindblad, \textit{Commun. Math. Phys.} \textbf{48}, 119 (1976).

\bibitem{30} 
A. Kossakowski, \textit{Rep. Math. Phys.} \textbf{3}, 247 (1972);  \\
V. Gorini, A. Kossakowski, and E. C. G. Sudarshan, \textit{J. Math. Phys.} \textbf{17}, 821 (1976).

\bibitem{31} 
F. Bloch, \textit{Phys. Rev.} \textbf{70}, 460 (1946).

\bibitem{32} 
R. Feynman, F. Vernon, and R. Hellwarth, \textit{Jour. of Appl. Phys.} \textbf{28}, 49 (1957).

\bibitem{33} 
G. Feinberg and S. Weinberg, \textit{Phys. Rev.} \textbf{123}, 1439 (1961).

\bibitem{34} 
S. V. Demidov, D. S. Gorbunov, and A. A. Tokareva, \textit{Phys. Rev.} \textbf{D 85}, 015022 (2012), \texttt{arXiv:1111.1072 [hep-ph]}.



\bibitem{35}
A.P. Serebrov, E.B. Aleksandrov, N.A. Dovator, et al., \textit{Nucl. Instr. and Methods} \textbf{611}, 137 (2009), \texttt{arXiv:0809.4902 [nucl-ex]}.

\bibitem{36}
A.P. Serebrov, E.B. Aleksandrov, N.A. Dovator, et al., \textit{Phys. Lett.} \textbf{B 663}, 181 (2008), \texttt{arXiv:0706.3600 [nucl-ex]}.

\bibitem{37}
Z. Berezhiani, R. Biondi, P. Geltenbort, et al., \textit{Eur. Phys. J.} \textbf{C 78}, 717 (2018), \texttt{arXiv:1712.05761 [hep-ex]}.

\bibitem{38}
C. Abel, N.J. Ayres, G. Ban, et al., \textit{Phys. Lett.} \textbf{B 812}, 135993 (2021), \texttt{arXiv:2009.11046 [hep-ph]}.

\bibitem{39}
F.M. Gonzalez, C. Rock, L.J. Broussard, et al., \textit{Phys. Rev.} \textbf{D 110}, 072022 (2024), \texttt{arXiv:2402.15981 [hep-ex]}.

\bibitem{40}
Yu. Kamyshkov, J. Ternullo, L. Varriano, et al., \textit{Symmetry} \textbf{14}, 230 (2022), \texttt{arXiv:2111.01791 [hep-ph]}.

\bibitem{41} 
P. Pearle, \textit{Eur. J. Phys.} \textbf{33}, 805 (2012), \texttt{arXiv:1204.2016 [math-ph]}.

\bibitem{42}
S. Weinberg, \textit{Phys. Rev.} \textbf{A 94}, 042117 (2016), \texttt{arXiv:1610.02537 [quant-ph]}.

\bibitem{43}
E. Braaten, H.-W. Hammer, and G. P. Lepage, \textit{Phys. Rev.} \textbf{A 95}, 012708 (2017), \texttt{arXiv:1607.08084 [cond-mat.quant-gas]}.

\bibitem{44}
A. D. Dolgov, \textit{Phys. Rept.} \textbf{370}, 333 (2002), \texttt{arXiv:hep-ph/0202122}.

\bibitem{45}
Y. Akamatsu, \textit{Phys. Rev.} \textbf{D 91}, 056002 (2015), \texttt{arXiv:1403.5783 [hep-ph]}.

\bibitem{46}
N. Brambilla, N. Brambilla, M. A. Escobedo, \textit{et al.}, \textit{Phys. Rev.} \textbf{D 108}, L011502 (2023), \texttt{arXiv:2302.11826 [hep-ph]}.

\bibitem{47}
H. Nikolic, \textit{JCAP} \textbf{04}, 002 (2015), \texttt{arXiv:1502.04324 [hep-th]}.

\bibitem{48}
R. N. Mohapatra and S. Nussinov, \textit{Phys. Lett.} \textbf{B 776}, 22 (2018), \texttt{arXiv:1709.01637 [hep-ph]}.

\bibitem{49}
P. S. Shternin, M. Baldo, and P. Haensel, \textit{Phys. Rev.} \textbf{C 88}, 065803 (2013), \texttt{arXiv:1311.4278 [astro-ph.SR]}.

\bibitem{50}
A. Gal, \textit{Phys. Rev.} \textbf{C 61}, 028201 (2000), \texttt{arXiv:hep-ph/9907334}.

\bibitem{51}
J. Redondo, \textit{JCAP} \textbf{07}, 024 (2015), \texttt{arXiv:1501.07292 [hep-ph]}.

\bibitem{52}
S. Demidov, S. Gninenko, and D. Gorbunov, \textit{JHEP} \textbf{07}, 162 (2019), \texttt{arXiv:1812.02719 [hep-ph]}.

\bibitem{53}
Z. Berezhiani, \textit{Universe} \textbf{8}, 313 (2022), \texttt{arXiv:2106.11203 [astro-ph.HE]}.



\end{thebibliography}
\end{document}